%% file: Manuscript.tex
\documentclass[graybox]{svmult}
\usepackage{type1cm}
\usepackage{makeidx}
\usepackage{graphicx}
\usepackage{multicol}
\usepackage[bottom]{footmisc}
\usepackage{newtxtext}
\usepackage{newtxmath}
\makeindex
\usepackage[ruled,vlined,linesnumbered]{algorithm2e}
\usepackage{caption}
\usepackage{url}

\begin{document}

\title*{Graph based Clustering Algorithm for Social Community Transmission Prediction of COVID-19}
\titlerunning{Graph Clustering for Transmission Prediction of COVID-19}
\author{Varun Nagesh Jolly Behera, Ashish Ranjan and Motahar Reza}
\institute{Varun Nagesh Jolly Behera \at R\&D Division, Just Another Media Laboratory (JAM Lab), Mumbai
\newline\email{varun.behera@jamlab.in}
\and Ashish Ranjan \at R\&D Division, Just Another Media Laboratory (JAM Lab), Mumbai
\newline\email{ashish.ranjan@jamlab.in}
\and Motahar Reza \at Department of Mathematics, GITAM University Hyderabad Campus, Hyderabad
\newline\email{mreza@gitam.edu}}
%
%
\maketitle

\abstract{A system to model the spread of COVID-19 cases after lockdown has been proposed in this paper, It helps to define preventive measures based on predicted hotspots, using the graph clustering algorithm. This method allows for more lenient measures in areas less prone to the virus spread. There exist methods to model the spread of the virus, by predicting the number of confirmed cases. But the proposed system focuses more on the preventive side of the solution from a geographical point of view. The fact that the virus can only be transmitted by being in close proximity to an already infected person, suggests that, the regions that can easily be reached from an existing hotspot, have a higher chance of becoming a new hotspot. Moreover, in smaller regions, even after strict provisions, positive cases have been found. So, the geodesic between the nearest hotspots can be used as a measure of likelihood of the region also becoming a hotspot. In this paper, a weighted graph of regions, with weight of the nodes as the number of active cases, and the distance as edge weights is used. The graph is connected based on a distance threshold. The nodes are the administrative level, and the distance measure tells the possible transmission between separate communities. Using this data, the potential regions that can become hotspots can be predicted, and preventive measures can be devised.}


\section{Introduction}
\label{sec:1}
The 2020 pandemic, caused by the SARS-CoV-2 \cite{jiang2020distinct} \cite{mahase2020covid}, popularly known as COVID-19 has resulted in thousands of deaths and millions being infected. It was officially first identified in December 2019 in Wuhan, China \cite{zhou2020pheumonia}. In mid March the World Health Organization declared it a pandemic, changing it's previous designation as an outbreak. Over 25 million have been infected and more than 800,000 dead worldwide, as of 6th September 2020 \cite{jhu2020dash}. 

Similar to most flu viruses, it is spread via sneeze and cough droplets \cite{feng2020rational}. Touching the face after touching a contaminated surface may also cause infection. Those who are asymptomatic may also spread the disease, although chances of infection spreading this way are low. The virus is most contagious starting a few days before the initial symptoms, and then staying the same a week or two later \cite{lauer2020incubation} \cite{lipsitch2020defining}.

Cough, fever, tiredness, breathing issues, are a few common symptoms. Since this virus primarily causes pneumonia and respiratory issues, those with existing conditions are more at risk. This is evident by the fact that almost 80\% deaths in China were aged above 60 \cite{zhou2020clinical}, most having pre-existing health complications like diabetes and heart diseases. The ratio of number of death to number of cases is 3.3\%, as of 6th September 2020, according John Hopkins university.

Social Distancing has been one of the best preventive measure for COVID-19, accompanied by wearing face masks, maintaining personal hygiene, regularly sanitizing one's surroundings, and self-isolating if flu like symptoms evolve \cite{adhikari2020epidimiology}. Governments have imposed lockdowns, travel-restrictions and mandatory safety measures to control the virus spread \cite{lau2020positive}. 

While the above measures prove to be very effective, they are not 100\% foolproof. Restrictions should be based on local situations, rather than be imposed at a country level. It makes no sense to restrict movement in an area that has next to none cases \cite{alvarez2020simple}. A balance is required to make sure the economy does not completely destabilise, and people not end up in a position where they are so financially unstable that they may starve to death. At the same time people need to protected from the infection too. A developing country like India will end up suffering from both a dying economy, and a dying workforce, if dynamic real time decisions are not taken \cite{baker2020covid}.

Most research done regarding disease epidemiology is done on factors like total case numbers, infection rates, and primarily focus on the SEIR (susceptible - exposed - infected - recovered) model \cite{yang2020modified}. But COVID-19 is unique in the sense that after imposing travel restrictions on air and rail, the only way transmission is possible via roads. This means the virus does not have much spatial transmission range, once mass travel restrictions and lockdowns are imposed. This makes spatial epidemiology  an ideal candidate for analyzing and predicting the spread of this type of disease.

A possible model for analyzing diseases like COVID-19, post-lockdown stage, is visualizing administrative zones as graph nodes, and the roads connecting them as the edges. This methodology will work regardless of the administrative level to be analyzed, as long as the zone-wise infection data is available, and has been accurately documented.

\section{Related Work}
\label{sec:2}
Epidemiology is generally either observational or experimental. Experimental studies follow a controlled factor based approach, while observational studies go for an entirely natural approach, with no interference  \cite{stallybrass1931principles}. Epidemiological studies should reveal relations with underlying health conditions. For example, it is known that cardiovascular diseases, diabetes, and other conditions, severely affect a COVID-19 patient. This paper focuses on observational studies, specifically on spatial epidemiology \cite{elliot2000spatial}.

A lot of work has been done on spatial epidemiology research. One of them being the mapping of diseases, allowing for better understanding of disease spread, and assisting the government in making well informed decisions on resource allocation \cite{ostfield2005spatial}. There also exist methodologies, based on the geographical demographics, including factors like race, age, etc that may influence the spread of diseases. \cite{kirby2017spatial}. Disease clusters, infection groups, hotspots, are also a dynamic to be analyzed. Although there are some issues to these approaches, provided the study has been done properly, they can be beneficial, especially for epidemics and pandemics \cite{beale2008methodologic}.

There has also been a lot of recent research specific to COVID-19. More prominent studies are based in China, being the origin point of the virus. For example a study \cite{kang2020spatial}, explores the spatial epidemic dynamics of COVID-19 using Moran's I spatial statistic. This study was based in Mainland China. It concluded that in fact, there is a spatial correlation regarding spread of COVID-19. A study specific to the difference is spatial transmission effects due to different public transport methods also exists, again based in China \cite{zheng2020spatial}. Most studies are based on the SEIR model \cite{huang2020spatial}, although SEIRS \cite{cooke1996analysis} should have been the way to go, due to WHO stating that reinfection could be possible. Other recent work includes using machine learning, cloud computing and IOT for modelling the COVID-19 trend \cite{tuli2020predict}, estimation of the reproduction number of 2019-ncov \cite{zhao2020pre}, and clustering analysis of countries for COVID-19 \cite{zarikas2020cluster}.

Studies based on the Indian scenario are more in line with the work explained in this paper. There have been a few studies with regards to spatial transmission, incuding multi-city modelling \cite{pujari2020multi}, trend analysis using a geospatial approach \cite{murugeshan2020distribution}, and spatial prediction using ARIMA \cite{roy2020spatial}.

\section{Goals and Objectives}
\label{sec:3}
The objective of this paper is to the study the spatial epidemiology of diseases like COVID-19, and devise a methodology to do so. This methodology will help predict future disease hotspots, and help in devising preventive measures as necessary. The modelling and prediction will be done at district level in this paper, but the same methodology can be applied on any administrative level. Following are a few goals of this paper:
\begin{enumerate}
    \item{\textbf{Country Independent}: The methodology needs to be independent of the country. As long as the assumptions are met, the algorithm should work automatically for any and all countries.}
    \item{\textbf{Administrative Level Independent}: It also should be independent of the administration level that needs to be analyzed. This includes state, county/district, zone, zip codes, etc. As long as both location and infection data is available, the algorithm should run flawlessly.}
    \item{\textbf{Automatic Graph Construction}: Graph for that administrative level should be generated automatically on the fly. This includes connecting nodes using clustering, forming edges, and assigning weight to both the nodes and edges.}
    \item{\textbf{Division Inclusive/Exclusive}: Impact of a location on itself may be included or excluded, based on user preference. This will give slightly different result in each case.}
    \item{\textbf{Real Time Updates}: The system should update periodically, preferably daily, so that better strategies can made for disease spread prevention and control. The result can be visualized by creating a web app dashboard.}
\end{enumerate}

\section{Requirements and Assumptions}
\label{sec:4}
The algorithm defined is this paper is robust, as long as few base assumptions are met. The reliability of this method is based on a lot of requirements and assumptions including:
\begin{enumerate}
    \item{\textbf{Restricted Travel}: This means that both air and rail travel are restricted. Also long distance travel by road is strictly monitored, making it difficult for the disease to have a large radius of influence.}
    \item{\textbf{Data Quality}: The data needs to be well defined, and properly cleaned. If the geographical names do not match with the Map API, then the longitude and latitude information of those locations cannot be retrieved. This means it is necessary to make sure there are no naming discrepancies, and the data should be modified as required, so that it's information aligns with that of the Map API to be used. Not checking and refining the data will result in varied amounts of information loss.}
    \item{\textbf{Data Granularity}: This model can work on various administrative levels including at the district, state and country level. Modelling can be done for a specific political map level only if the data is available for it. Meaning, if the model is to be defined at district level, both the Map API and the disease spread data should have granularity till the district level. If, either the co-ordinate data or the disease spread data of districts are missing, this model cannot be applied at that level.}
    \item{\textbf{Data Accuracy}: The data needs to be accurate. If data is old or incorrect, the predictions made by the model will be drastically affected. Accuracy of the current number of infected is most important. This means testing more, and using better quality tests will help this model a lot.}
    \item{\textbf{Proper Map API}: Map API that gives accurate co-ordinates of the geographical locations at the required administrative level is necessary. Only then can the graph be created with proper weight values and connections.}
\end{enumerate}

\section{Methodology Overview}
\label{sec:5}
\begin{figure}
\centering
\captionsetup{justification=centering}
\includegraphics[width=\textwidth,height=\textheight,keepaspectratio]{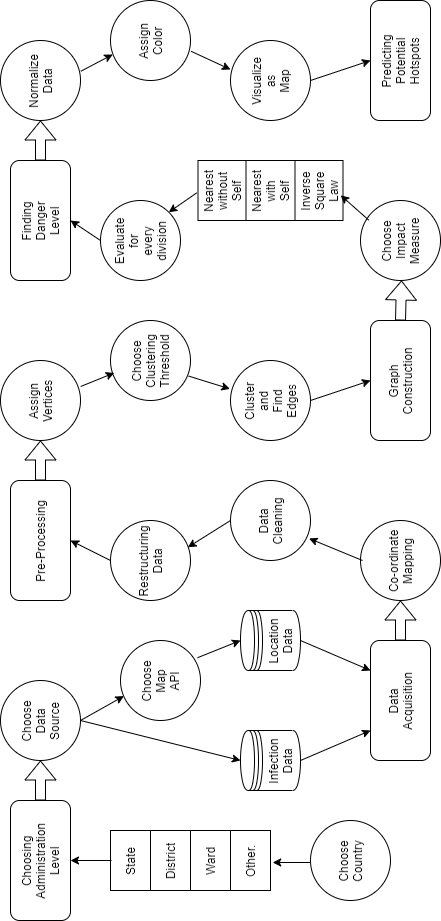}
\caption{Methodology Overview}
\label{fig:1}
\end{figure}

Following is a brief overview of the workflow described in this paper. It is visually represented in Fig.\ref{fig:1}.
\begin{enumerate}
    \item{\textbf{Choosing Administrative Level}: The first step is choosing the administrative level, at which infection spread has to be analyzed and predicted. Administrative levels including, but not limited to, countries, states, counties, districts, zones, wards, zipcodes, can be chosen for analysis. The only limitation in choosing the administrative level, is the unavailability of data at that level. Otherwise, any and all administrative levels can be chosen and analyzed.}
    \item{\textbf{Data Acquisition}: The reliability of the algorithm is entirely based on the 'goodness' of data. This includes aspects like availability, accuracy, comprehensiveness, etc. If both, infection data and location data are available and reliable, only then can the algorithm be used. The algorithm does not have the ability to fix inaccuracies and errors in the data. It is assumed that the data is accurate enough to be used. There exist many infection data sources, depending on the disease whose transmission has to be studied. For the purposes of finding location data, a satisfactorily capable Map API is used.}
    \item{\textbf{Pre-processing}: The data obtained from the previous step has to be cleaned and restructured properly. This includes dropping unnecessary columns from the infection data, and infection count that has not been assigned to any administrative division. It is also necessary to make sure that the latitudes and longitudes are properly mapped with their respective divisions.}
    \item{\textbf{Graph Construction}: Based on the restructured data, individual divisions are designated as separate graph nodes. This is done to create an abstract view of the countries administrative divisions, and how they are connected. Using a clustering method inspired from k-means, nearest nodes for every node are evaluated. All pairs of nearest nodes are connected to form graph edges. A distance metric is used to define the clustering factor. Only divisions within a certain range will be connected to each other. This distance metric has to be changed based on the administrative level chosen, to factor in the difference in distances between division centres.}
    \item{\textbf{Finding Danger Level}: Three different measures have been defined in this paper, to evaluate the probability of a district becoming a hotspot, termed as 'danger level'. These measures evaluate this danger level, using different parameters, giving different levels of importance to each parameter.}
    \item{\textbf{Predicting Potential Hotspots}: The danger level value obtained can be easily color coded. The color green shall represent a division having low risk of becoming a hotspot anytime soon. Orange signifies that the division is near a few existing hotspots, and if precautions are not taken, may become one itself. Red marks regions that either already have, or will very soon become major hotspots, unless strict measures are not taken.}
\end{enumerate}

\section{Data Acquisition}
\label{sec:6}
This model requires primarily two types of datasets. First is information regarding the current number of infections at the desired administrative level. This paper explores the models robustness at district level, but as mentioned before, this model can be used at any administrative level. Second is the information regarding the co-ordinates of various locations at same administrative level. This information can be extracted efficiently using a Map API.

\subsection{Infection Data}
\begin{figure}
\centering
\captionsetup{justification=centering}
\includegraphics[width=\textwidth,height=\textheight,keepaspectratio]{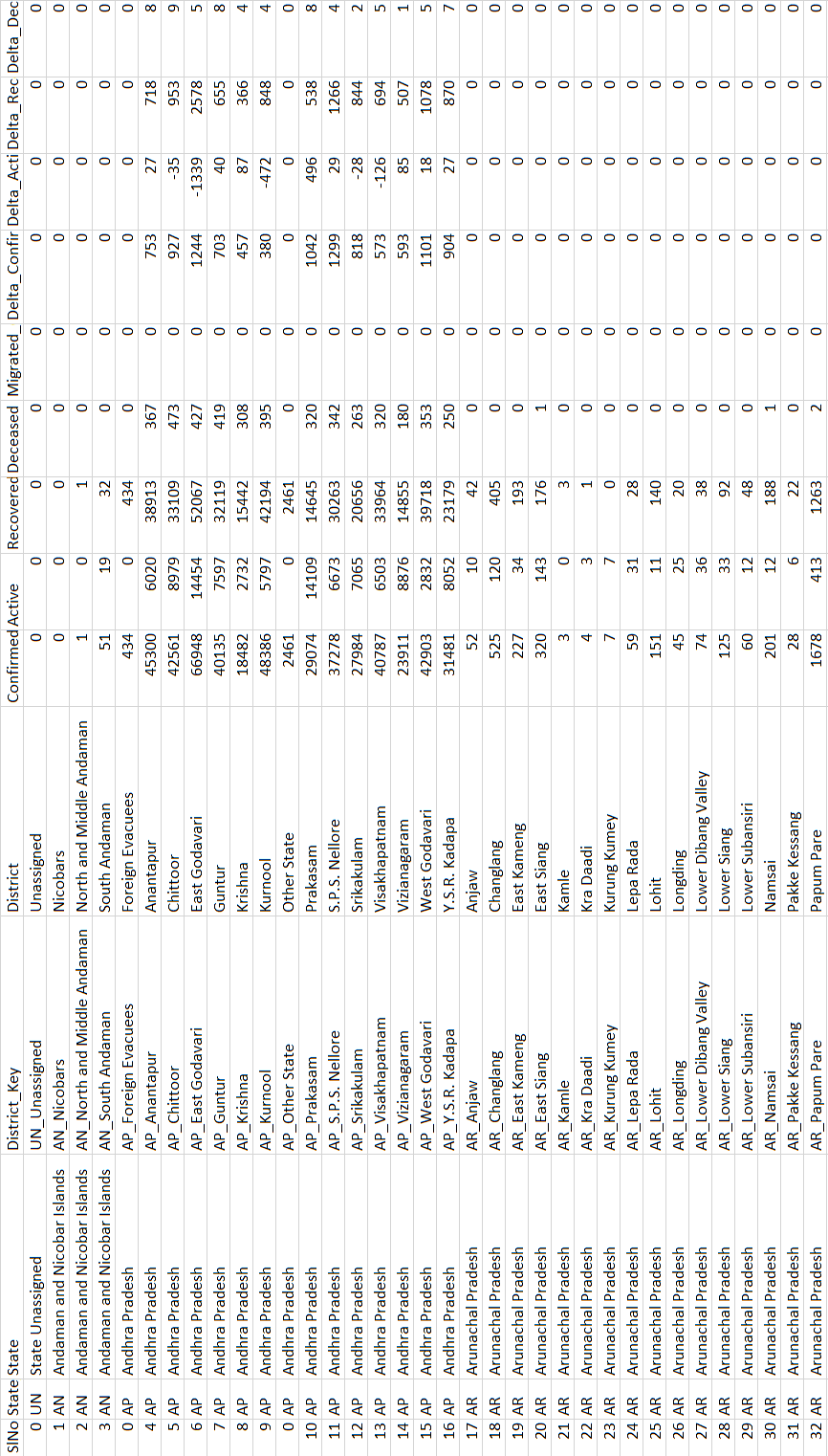}
\caption{Raw Infection Data}
\label{fig:2}
\end{figure}

This is the data of the total number of active infections, at the required administration level. Varied number of sources exist, depending on the disease that needs to be studied. Some prominent datasets include the Malaria Atlas Project \cite{hay2006malaria}, and Scientific Data for Dengue \cite{messina2014global}, Leishmania \cite{pigott2014global} and Ebola \cite{mylne2014comprehensive}. This paper explores the impact of COVID-19 in the Indian scenario. Some datasets currently available for it are nCoV-2019 \cite{xu2020epidimiological}, CORD-19 \cite{wang2020covid}, Lancet Dashboard \cite{dong2020interactive} and Coronavirus Twitter Dataset \cite{chen2020first}. Since this paper explores the Indian scenario at district level, the dataset chosen is from covid19india.org \cite{covid2020india}, a crowd-sourced dataset,  as seen in Fig.\ref{fig:2}. This dataset is highly accurate and is updated daily, sourced both from official government sources and media. There also exists an official Indian Government Dataset that may be used for state level analysis.

\subsection{Location Data}
\begin{figure}
\centering
\captionsetup{justification=centering}
\includegraphics[width=\textwidth,height=\textheight,keepaspectratio]{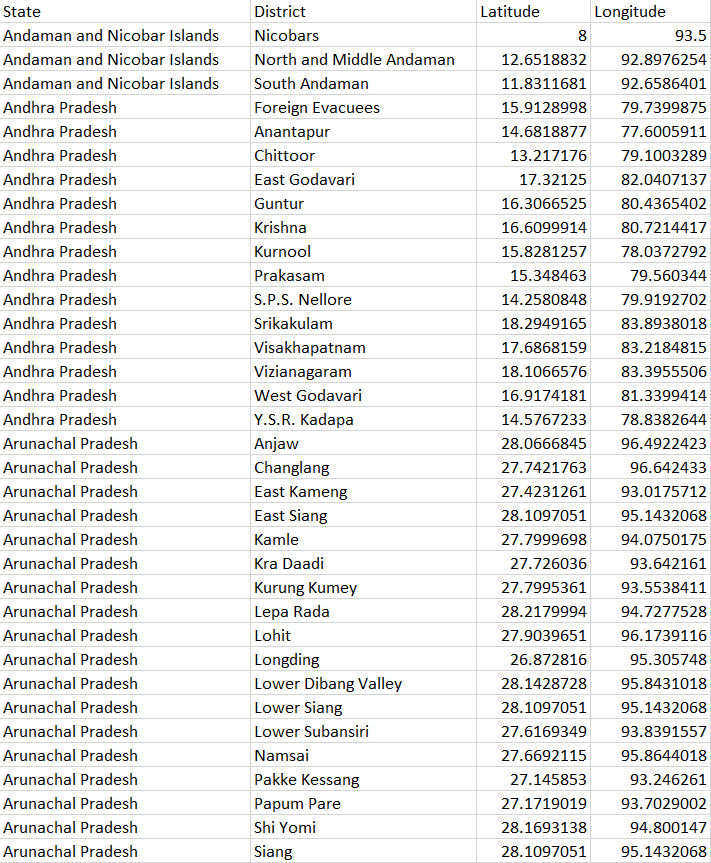}
\caption{Location Data}
\label{fig:3}
\end{figure}

This is the data of the administrative division location at the required administrative level. This data can be obtained using a Map API, in the form of latitudes and longitudes. The Map API will take the location string as input and give co-ordinates as output, as shown in Algo. \ref{algo:1} and Fig.\ref{fig:3}. For the case of districts, the upper level will be state. The most popular Map APIs are the Bing Maps API and the Google Maps Geocoding API. 

\begin{algorithm}
\DontPrintSemicolon
\SetAlgoLined
\SetKwInOut{Input}{Input }
\Input{Location List}
\KwResult{Location with corresponding Co-ordinates}
\BlankLine
\BlankLine
READ locationList\;
APPLY Preprocessing\;
\ForEach{location in administrativeLevel}{    
    \If{location != 'Unassignedd' or 'Unknown' or 'Other'}{
        coordinates = geocode(location, upperLevel(location))\;
        APPEND location, coordinates to list L\;
    }
}
\caption{Algorithm for finding co-ordinates of every location}
\label{algo:1}
\end{algorithm}

\subsubsection{Bing Maps API}
There are two ways to find longitudes and latitudes using the Bing Maps API. 
        
First is using the 'Find location by address method'. In this method we use a specific URL template to obtain longitudes and latitudes. This is done by filling various parameters in the template URL, like locality, postal code, and street address. To use the API a Bing Maps key is required. Requesting a location using the URL template returns a response containing location data like latitudes and longitudes, location type, and the geographical area that contains the location. The API may be accessed using both HTTP and HTTPS protocols. Information for a location can be obtained by setting either one or more parameters. The output can be obtained either as an XML or JSON response. Both responses will give central latitude and longitude, as well as latitudes and longitudes of the bounding box for that location.
        
The second is using the 'Find a location by query' method. This method gives longitudes and latitudes, based on a given query. This method is more akin to generally how Map Applications are interacted with generally. The query can be sent as a structured URL or as an unstructured query string. User Context Parameters  may be used for better accuracy. The output is similar to the previous method, although more than one output may be received due to ambiguity. It also likewise supports both HTTP and HTTPS.

\subsubsection{Google Geocoding}
A part of the Google Maps Platform, subset to the Google Cloud Platform, this API is similar to the second Bing Maps API method. This API service can be directly accessed using an HTTP request. An API key has to be obtained from the Google Cloud Platform to use the Geocoding API. Google's API also gives output in JSON or XML as needed. Input can be either a string, containing the address, or as components with their corresponding values. The output will include both, the central latitude and longitude and the bounding box latitudes and longitudes. The benefit of using Google Geocoding is the support for upto 5 administrative levels within a country. Districts are at the administrative area level 2 as per the API.

\section{Pre-processing}
\label{sec:7}
\begin{figure}
\centering
\captionsetup{justification=centering}
\includegraphics[width=\textwidth,height=\textheight,keepaspectratio]{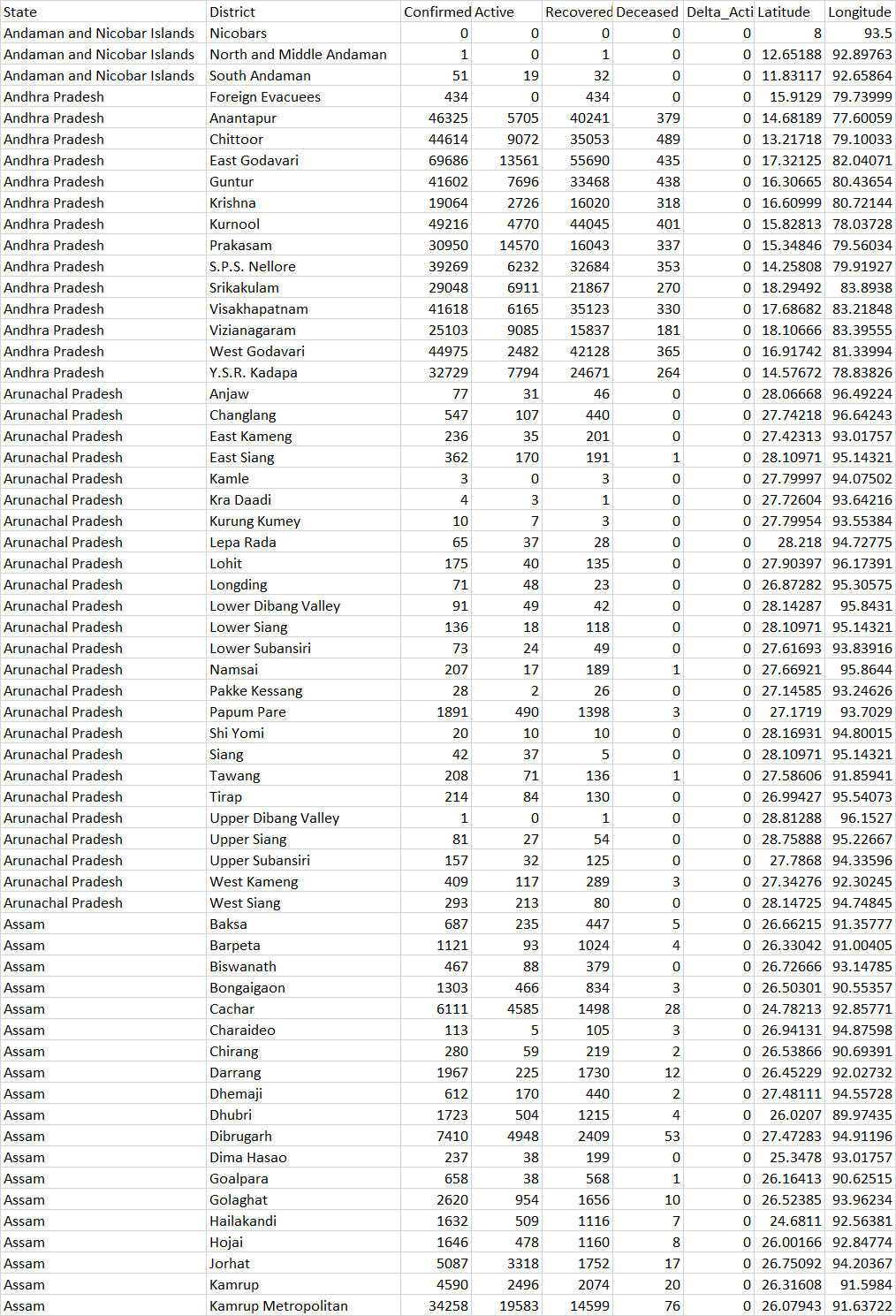}
\caption{Final Data}
\label{fig:4}
\end{figure}

The model requires administrative division location with its corresponding active number of cases. Before this can be done, co-ordinates i.e. latitudes and longitudes for every division has to be found. The infection data too needs to be cleaned. This includes dropping unnecessary columns and removing rows with undefined divisions. After that both datasets need to be mapped to give the required restructured data. Since this paper focuses on the district level, mapping will be done at that level, as shown in Fig.\ref{fig:4}.

\subsection{Co-ordinate Mapping}
This paper explores using the Google Geocoding API for finding latitudes and longitudes of administrative divisions at the required level. The 'googlemaps' package is used, accompanied with the Geocoding API client key. For every district-state pair in the infection data, a query is formed as 'District, State'. For each such query, a corresponding latitude-longitude pair is obtained. All of these obtained location-coordinate pairs are properly stored.

\subsection{Data Cleaning}
The data obtained from covid19india.org contains a lot of information, most of which is not required for the study done in this paper. This includes attributes like State Code, Confirmed cases, Recovered cases, etc. The attributes required are State, District, Active cases, and Change in Active cases. Change in Active cases is the increase in the number of active cases from the previous day. This is because a district with higher number of currently active cases will have a more severe impact on its neighbouring districts. The use of each of these parameters are mentioned later in this paper.

\subsection{Restructuring Data}
After data has been cleaned, the geolocation data obtained from the Map API is merged with this cleaned data. The final data will contain District, State, Active Cases, Change in Active Cases, Latitude and Longitude values for every District.

\section{Graph Construction}
\label{sec:8}
\begin{figure}
\centering
\captionsetup{justification=centering}
\includegraphics[width=\textwidth,height=\textheight,keepaspectratio]{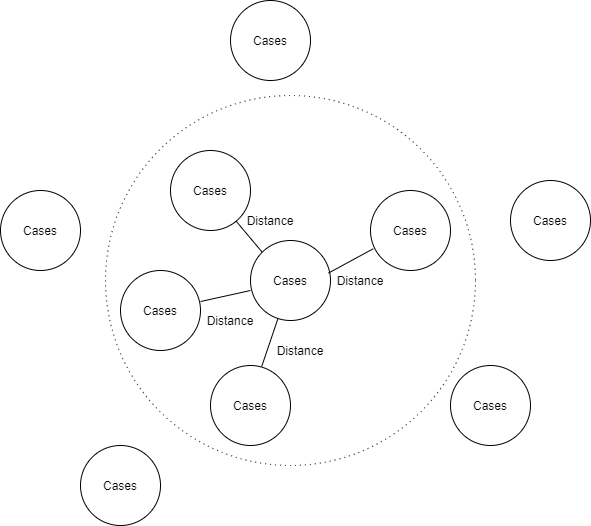}
\caption{Sample Graph}
\label{fig:5}
\end{figure}

The next step is to construct the graph using the newly restructured data. Each of the divisions/districts shall be considered as a vertex, and few of its connections will be edges. The main aim of creating a graph is to form a high-level view of a country's transport system, or specifically, it's road connections. So, the vertices are the districts, and the edges are the shortest paths between neighbouring districts, as shown in Fig.\ref{fig:5} and Algo.\ref{algo:2}. Cases are the vertex weights, shortest distances are the edge weights, and the dotted line represents the threshold boundary. This entire process is accomplished using 'networkx' package.

\subsection{Vertices}
For constructing the graph, the district locations are considered as the vertices. The weight for each vertex is the corresponding number of active cases. If any other administrative level is to be considered, the respective division locations are considered as vertices. Accordingly the vertex weight for each division will be number of active cases for that division.

\subsection{Edges}
Edges for the graph are the connections between only some specific pairs of vertices. This is because, districts nearby each other are easier to travel to and from, due to air and rail travel restrictions. Also, even if all pairs are considered, the impact from a district very far away, even with high number of cases, will be negligible. It will only result cause more computational complexity. These pairs can be obtained by using some clustering methodology. After that, the weight for an edge can be defined by the distance between its two vertices.

\subsubsection{Clustering} 
The simplest and most logical approach to find appropriate pairs is to take inspiration from something like k-means of fuzzy c-means clustering \cite{jain2010data}. Basically, for every vertex in the graph, if another vertex is within some threshold range of that vertex, then they are connected. Each vertex is its own cluster centre, and cluster centres are not updated. Nearest vertices are identified in a single pass, and a vertex can be in multiple clusters. Every administrative division becomes its own cluster centre.

\subsubsection{Threshold Value}
Two vertices should only be connected based on the above mentioned threshold value. The threshold value is actually the maximum shortest distance between two districts, so that they may be considered as neighbours. This threshold distance needs to be appropriate and has to be set manually. A reasonably large value is chosen for threshold, for a balanced number of connections between districts.

\begin{algorithm}
\DontPrintSemicolon
\SetAlgoLined
\SetKwInOut{Input}{Input }
\Input{Final Data, Threshold}
\KwResult{Connected Graph}
\BlankLine
\BlankLine
READ Final Data, Threshold\;
\ForEach{k in divisions}{
    G.addNode(k, cases(k))
}
\ForEach{i in G.nodes}{
    \ForEach{j in G.nodes}{
        \If{i != j and i,j not in edge}{
            \If{distance(i,j) <= threshold}{
                G.addEdge(i, j, distance(i, j))\;
            }
        }
    }
}
\caption{Algorithm for generating Graph}
\label{algo:2}
\end{algorithm}

\subsection{Visualization}
Once the edges are defined, and both nodes and edges have their weights assigned, the graph construction is complete. Now, it can be visualized in the graph, that the nodes are the administrative division centres, and the edges are the shortest paths between nearby divisions.

\begin{figure}
\centering
\captionsetup{justification=centering}
\includegraphics[width=\textwidth,height=\textheight,keepaspectratio]{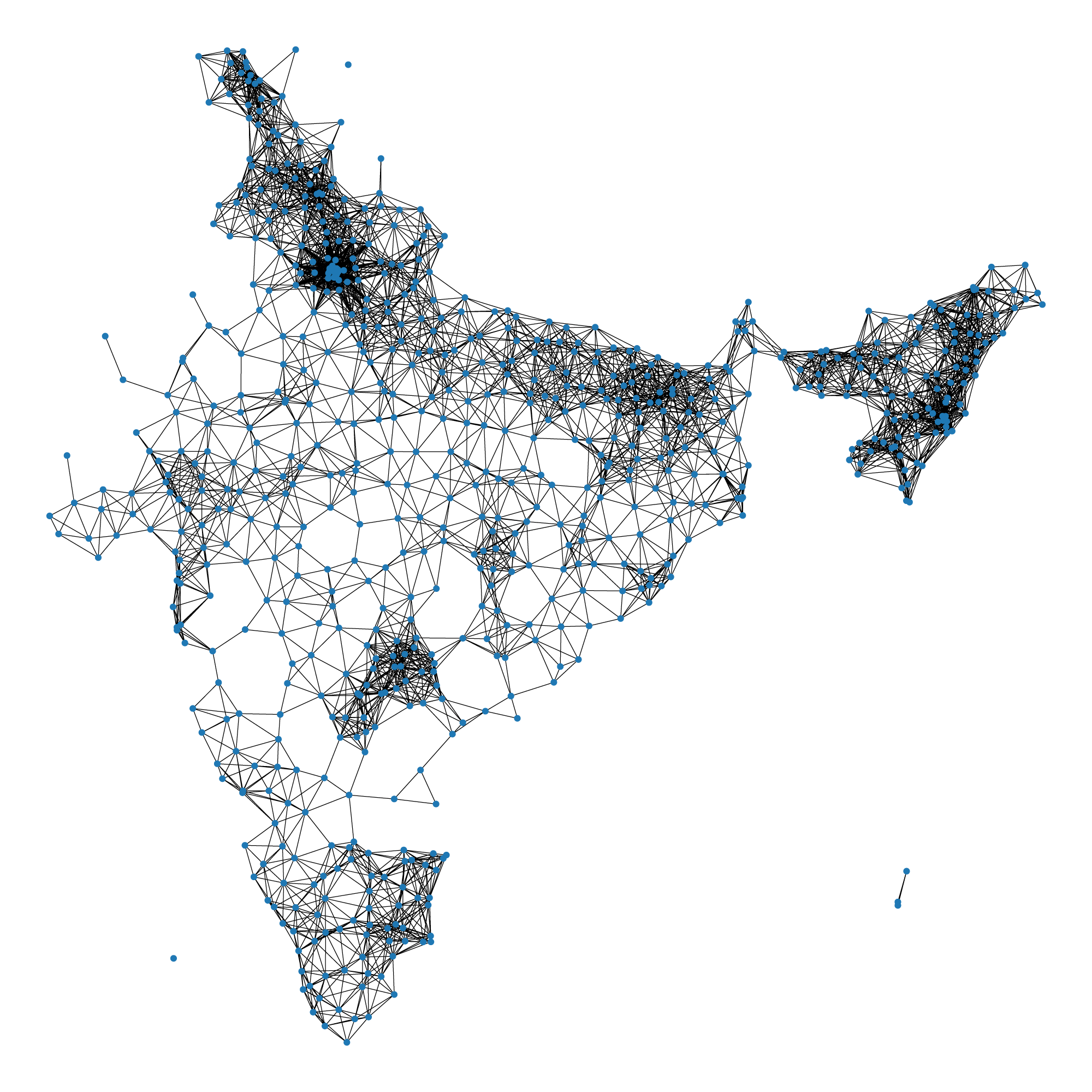}
\caption{Constructed Graph}
\label{fig:6}
\end{figure}

\section{Impact Measures and Danger Level}
\label{sec:9}
To find the danger level of each individual division, three measures have been defined. Each measure takes into account different factors, so results may be slightly different. These measures give a danger value, that may be color coded for better visualization.

\subsection{Nearest without Self}
In this measure, the impact of all other cluster members, except for the cluster centre itself, on the cluster centre, is taken  into consideration. This danger level is defined as

\begin{equation}
\label{eq:1}
    D_{NWOS}(i) =\frac{\sum_j^N impact(i,j)}{N}
\end{equation}

here $i$ identifies the division whose danger level needs to be evaluated, $j$ is any division adjacent to division $i$ i.e. $ij \in E$, and $N$ is the number of such pairs, where

\begin{equation}
\label{eq:2}
    impact(i,j) = \frac{V_{scaled}(j)}{E_{scaled}(i,j)}
\end{equation}

is the impact factor of $j$ division on $i$, and the scaled vertex weights and edge weights are

\begin{equation}
\label{eq:3}
    V_{scaled}(j)=V(j)/max(V)
\end{equation}

\begin{equation}
\label{eq:4}
    E_{scaled}(j)=E(j)/max(E)
\end{equation}

\subsection{Nearest with Self}
In this measure, the impact of all cluster members, including the cluster centre on itself, is taken  into consideration. This danger level is defined as

\begin{equation}
\label{eq:5}
    D_{NWS}(i) =\frac{\sum_j^N impact(i,j) + impact_{self}(i)}{N+1}
\end{equation}

here $i$ identifies the division whose danger level needs to be evaluated, $j$ is any division adjacent to division $i$ i.e. $ij \in E$, and $N$ is the number of such pairs, where $impact(i,j)$ is the same as Eq.\ref{eq:2}, and

\begin{equation}
\label{eq:6}
    impact_{self}(i)=V_{scaled}(i)\times(1+\triangle V(i)/V(i))
\end{equation}

is the impact factor of division $i$ on itself, and the scaled vertex weights and edge weights are same as Eq.\ref{eq:3} and \ref{eq:4}. If $V(i)$ is zero, then self impact will also be zero.

\subsection{Inverse-square Law}
In this measure, the impact of all cluster members, including the cluster centre on itself, is taken  into consideration. It is similar to Coulomb's Law and Newton's Gravitational Law. This danger level is defined as

\begin{equation}
\label{eq:7}
    D_{ISL}(i) =\frac{\sum_j^N impact(i,j)}{N}
\end{equation}

here $i$ identifies the division whose danger level needs to be evaluated, $j$ is any division adjacent to division $i$ i.e. $ij \in E$, and $N$ is the number of such pairs, where

\begin{equation}
\label{eq:8}
    impact_{ISL}(i,j) = \frac{V_{scaled}(i) \times V_{scaled}(j)}{E^2_{scaled}(i,j)}
\end{equation}

is the impact factor of $j$ division on $i$, and the scaled vertex weights and edge weights are same as Eq.\ref{eq:3} and \ref{eq:4}.

\section{Predicting Potential Hotspots}
\label{sec:10}
Using any of the three measures, the danger level value can be easily color coded. This will allow for ease in visualisation, as the various colors will represent different danger levels, in a gradient manner. Three colours are used, namely green, orange and red. Here red represents the division that is most severely affected, or about to be severely affected, very soon. This division is either already a hotspot, or will become one in a few days. Orange represents divisions that either are moderately affected, or will soon be, if correct precautions are not taken. This is the best stage to impose restrictions, so as to make sure it does not become a red zone i.e. a future disease hotspot. An orange zone may also signify that some of the nearby divisions may already red zones, solidifying the need for major precautions. Finally, green represents a comparatively safe zone, where, as long as there are restrictions on travelling in and out of that zone, lenient rules may be applied, making sure that people's lives and their livelihood both are not affected.

Since a few highly infected areas may be deemed outliers, a normal distribution is instead used, considering only the second and third quartile. After this, according to danger values, colours are assigned, with the lower danger values getting assigned green, and higher ones as red. The color gradient then helps in easily understanding the spatial situation and dynamically choosing the rules based on the zone.

\section{Experimental Results}
\label{sec:11}
Data acquisition, pre-processing, graph construction, and the algorithm implementation, were all done in Python 3, using the Google Collaboratory platform. Experimental evaluation was done using an Intel Xeon 2vCPU @ 2.3GHz and more than 12 GB of RAM. Infection data was acquired from covid19india.org, while Location Data was obtained using the Google Maps Geocoding API.

\begin{figure}
\centering
\captionsetup{justification=centering}
\includegraphics[width=\textwidth,height=\textheight,keepaspectratio]{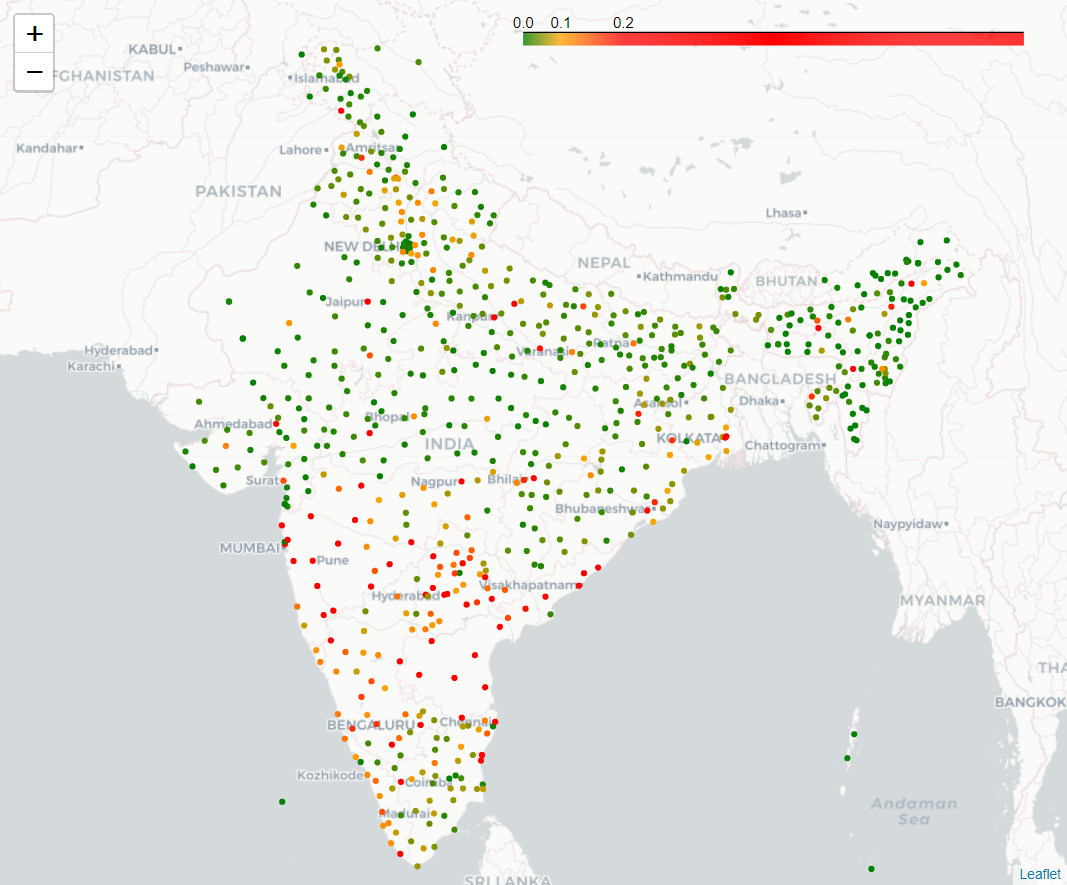}
\caption{Original Data}
\label{fig:7}
\end{figure}

\begin{figure}
\centering
\captionsetup{justification=centering}
\includegraphics[width=\textwidth,height=\textheight,keepaspectratio]{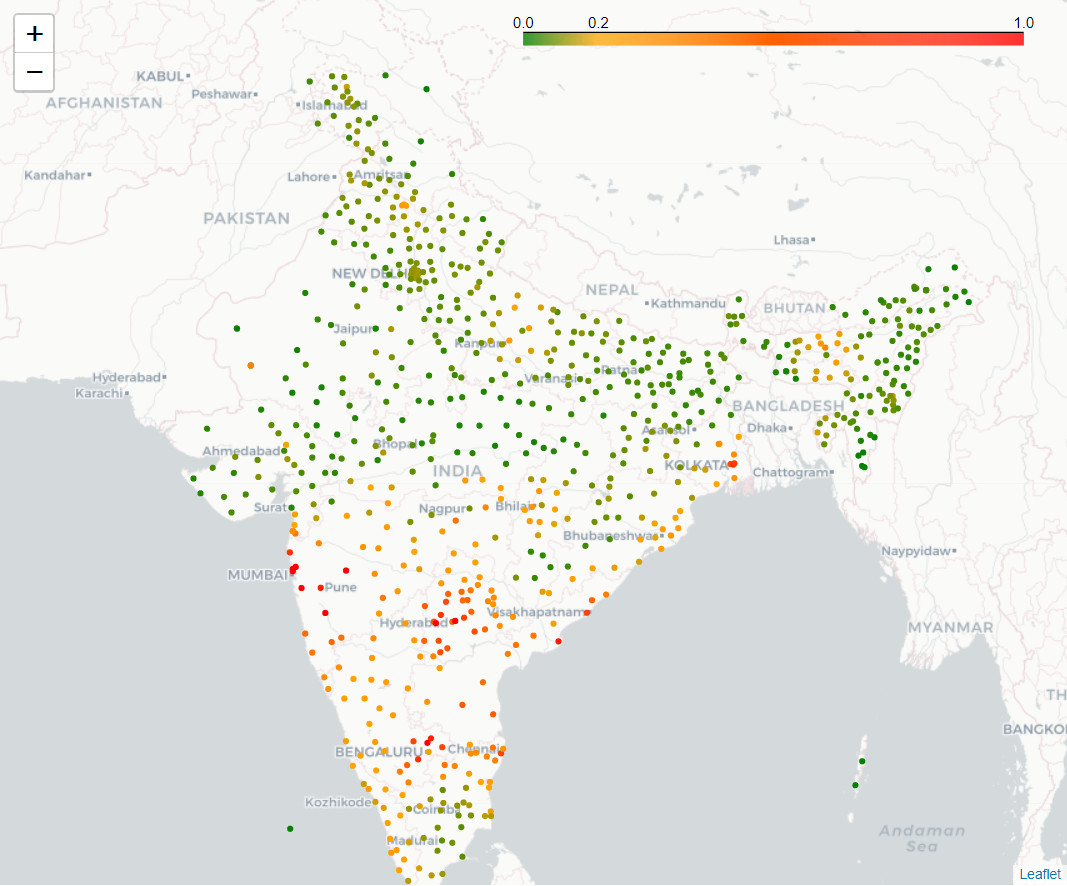}
\caption{Nearest Without Self}
\label{fig:8}
\end{figure}

\begin{figure}
\centering
\captionsetup{justification=centering}
\includegraphics[width=\textwidth,height=\textheight,keepaspectratio]{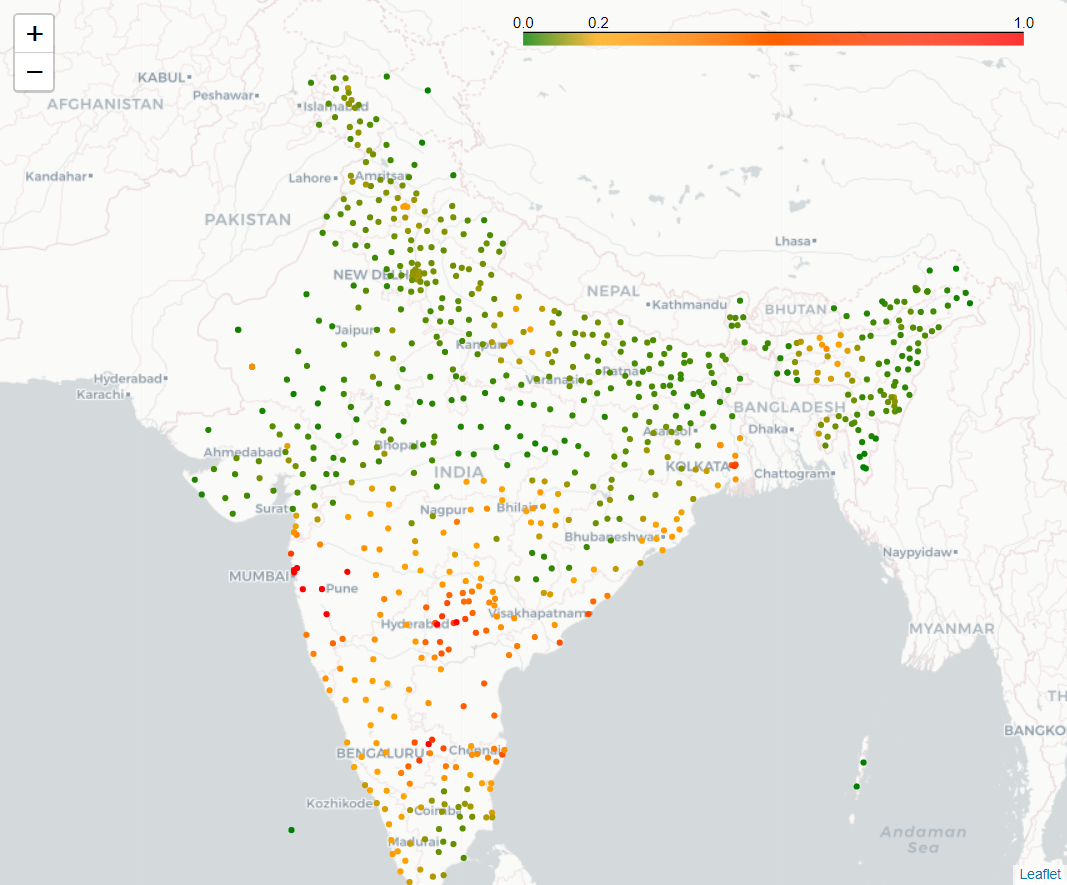}
\caption{Nearest With Self}
\label{fig:9}
\end{figure}

\begin{figure}
\centering
\captionsetup{justification=centering}
\includegraphics[width=\textwidth,height=\textheight,keepaspectratio]{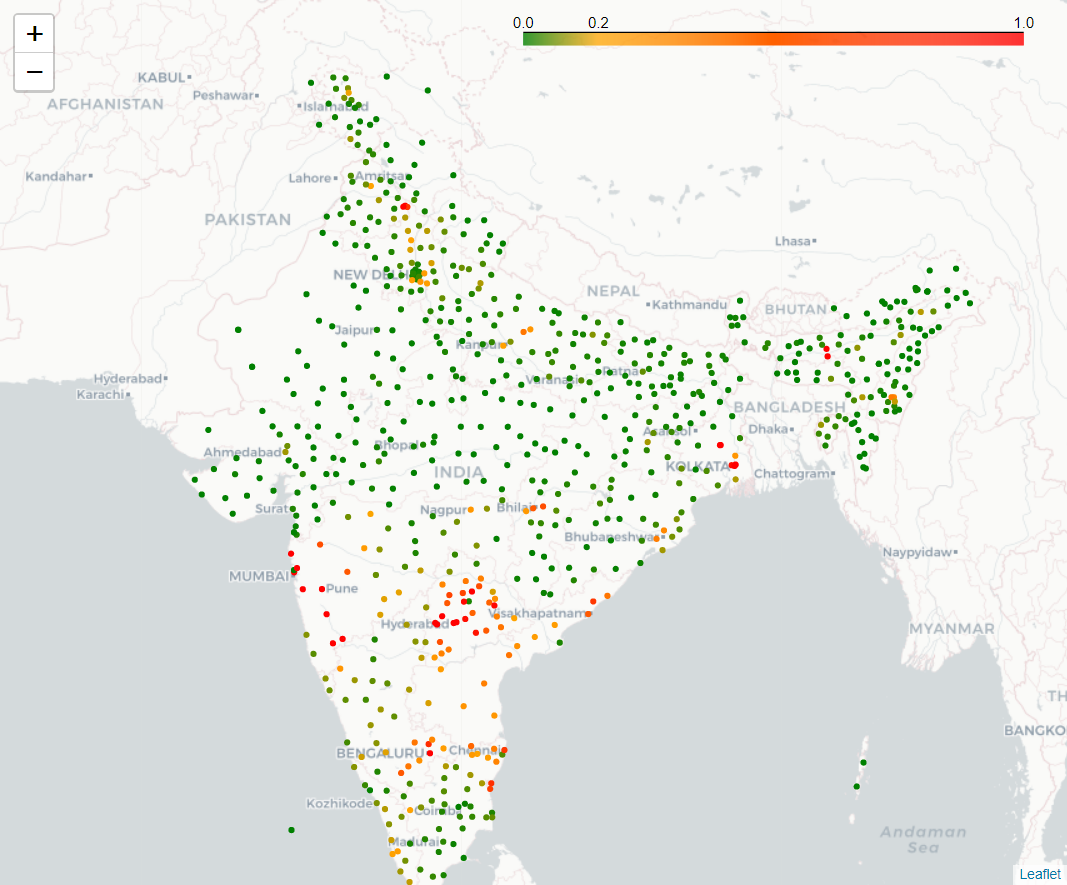}
\caption{Inverse Square Law}
\label{fig:10}
\end{figure}

With a threshold distance value of 1.3 degrees, a graph was generated (Visualized in Figures.\ref{fig:7}-\ref{fig:10}). This value was found to be large enough to connect all appropriate nodes while being computationally efficient.

To take care of very huge number of active cases that act as outliers, it was assumed that every node with a danger value of 0.3 and above be equal. The danger value was scaled to a value between 0 and 1. These values were mapped to a colour gradient map and the final map visualization of the country was generated. All three measures were used and compared.

The entire system has been converted to a real time web-app where the computation is scheduled daily and the final visualized map is shown on a dashboard. It is available at https://research.jamlab.in/covid .

\section{Conclusion}
\label{sec:12}
A system was proposed to model the spread of COVID-19 cases after lockdown, to define new preventive measures based on hotspots, using the graph clustering algorithm. The system works on any administrative level given that data is available at that level. A graph was constructed to model the virus' spread and danger values were calculated to assign the danger level of each district. This was visualised using a color map.

Although this system was tested on the COVID-19 India data, it should work on dataset for any disease, provided all assumptions that are mentioned are met. There is a lot of scope for this methodology, which will be explored in future work .

\section{Future Work}
\label{sec:13}
A lot of improvements can be made to this methodology. This includes addition of a time dimension, so as prediction can be made for a defined period of time. A pseudo geospatial time series analysis may be possibly used for this. Also instead of using a rough geodesic graph, an exact graph based n the actual road network can also be explored.

\input{references}

\end{document}

%% file: references.tex
%
%
%

%% file: Manuscript.bbl
\begin{thebibliography}{99.}%
%
%

\bibitem{jiang2020distinct} Jiang, Shibo, et al. "A distinct name is needed for the new coronavirus." Lancet (London, England) 395.10228 (2020): 949.

\bibitem{mahase2020covid} Mahase, Elisabeth. "Covid-19: WHO declares pandemic because of “alarming levels” of spread, severity, and inaction." (2020).

\bibitem{zhou2020pheumonia} Zhou, Peng, et al. "A pneumonia outbreak associated with a new coronavirus of probable bat origin." nature 579.7798 (2020): 270-273.

\bibitem{jhu2020dash} Johns Hopkins University. "COVID-19 Dashboard by the Center for Systems Science and Engineering (CSSE) at Johns Hopkins University (JHU)." (2020).

\bibitem{feng2020rational} Feng, Shuo, et al. "Rational use of face masks in the COVID-19 pandemic." The Lancet Respiratory Medicine 8.5 (2020): 434-436.

\bibitem{lipsitch2020defining} Lipsitch, Marc, David L. Swerdlow, and Lyn Finelli. "Defining the epidemiology of Covid-19—studies needed." New England journal of medicine 382.13 (2020): 1194-1196.

\bibitem{lauer2020incubation} Lauer, Stephen A., et al. "The incubation period of coronavirus disease 2019 (COVID-19) from publicly reported confirmed cases: estimation and application." Annals of internal medicine 172.9 (2020): 577-582.

\bibitem{zhou2020clinical} Zhou, Fei, et al. "Clinical course and risk factors for mortality of adult inpatients with COVID-19 in Wuhan, China: a retrospective cohort study." The lancet (2020).

\bibitem{adhikari2020epidimiology} Adhikari, Sasmita Poudel, et al. "Epidemiology, causes, clinical manifestation and diagnosis, prevention and control of coronavirus disease (COVID-19) during the early outbreak period: a scoping review." Infectious diseases of poverty 9.1 (2020): 1-12.

\bibitem{lau2020positive} Lau, Hien, et al. "The positive impact of lockdown in Wuhan on containing the COVID-19 outbreak in China." Journal of travel medicine 27.3 (2020): taaa037.

\bibitem{alvarez2020simple} Alvarez, Fernando E., David Argente, and Francesco Lippi. A simple planning problem for covid-19 lockdown. No. w26981. National Bureau of Economic Research, 2020.

\bibitem{baker2020covid} Baker, Scott R., et al. Covid-induced economic uncertainty. No. w26983. National Bureau of Economic Research, 2020.

\bibitem{yang2020modified} Yang, Zifeng, et al. "Modified SEIR and AI prediction of the epidemics trend of COVID-19 in China under public health interventions." Journal of Thoracic Disease 12.3 (2020): 165.

\bibitem{stallybrass1931principles} Stallybrass, Clare Oswald. "The Principles of Epidemiology and the Process of Infection." The Principles of Epidemiology and the Process of Infection. (1931).

\bibitem{elliot2000spatial} Elliot, Paul, et al. Spatial epidemiology: methods and applications. Oxford University Press, 2000.

\bibitem{ostfield2005spatial} Ostfeld, Richard S., Gregory E. Glass, and Felicia Keesing. "Spatial epidemiology: an emerging (or re-emerging) discipline." Trends in ecology \& evolution 20.6 (2005): 328-336.

\bibitem{kirby2017spatial} Kirby, Russell S., Eric Delmelle, and Jan M. Eberth. "Advances in spatial epidemiology and geographic information systems." Annals of epidemiology 27.1 (2017): 1-9.

\bibitem{beale2008methodologic} Beale, Linda, et al. "Methodologic issues and approaches to spatial epidemiology." Environmental health perspectives 116.8 (2008): 1105-1110.

\bibitem{kang2020spatial} Kang, Dayun, et al. "Spatial epidemic dynamics of the COVID-19 outbreak in China." International Journal of Infectious Diseases (2020).

\bibitem{zheng2020spatial} Zheng, Ruizhi, et al. "Spatial transmission of COVID-19 via public and private transportation in China." Travel Medicine and Infectious Disease (2020).

\bibitem{huang2020spatial} Huang, Rui, Miao Liu, and Yongmei Ding. "Spatial-temporal distribution of COVID-19 in China and its prediction: A data-driven modeling analysis." The Journal of Infection in Developing Countries 14.03 (2020): 246-253.

\bibitem{cooke1996analysis} Cooke, Kenneth L., and P. Van Den Driessche. "Analysis of an SEIRS epidemic model with two delays." Journal of Mathematical Biology 35.2 (1996): 240-260.

\bibitem{tuli2020predict} Tuli, Shreshth, et al. "Predicting the Growth and Trend of COVID-19 Pandemic using Machine Learning and Cloud Computing." Internet of Things (2020): 100222.

\bibitem{zhao2020pre} Zhao, Shi, et al. "Preliminary estimation of the basic reproduction number of novel coronavirus (2019-nCoV) in China, from 2019 to 2020: A data-driven analysis in the early phase of the outbreak." International journal of infectious diseases 92 (2020): 214-217.

\bibitem{zarikas2020cluster} Zarikas, Vasilios, et al. "Clustering analysis of countries using the COVID-19 cases dataset." Data in Brief (2020): 105787.

\bibitem{pujari2020multi} Pujari, Bhalchandra S., and Snehal M. Shekatkar. "Multi-city modeling of epidemics using spatial networks: Application to 2019-nCov (COVID-19) coronavirus in India." medRxiv (2020).

\bibitem{murugeshan2020distribution} Murugesan, Bagyaraj, et al. "Distribution and Trend Analysis of COVID-19 in India: Geospatial Approach." Journal of Geographical Studies 4.1 (2020): 1-9.

\bibitem{roy2020spatial} Roy, Santanu, Gouri Sankar Bhunia, and Pravat Kumar Shit. "Spatial prediction of COVID-19 epidemic using ARIMA techniques in India." Modeling Earth Systems and Environment (2020): 1-7


\bibitem{hay2006malaria} Hay, Simon I., and Robert W. Snow. "The Malaria Atlas Project: developing global maps of malaria risk." PLoS Med 3.12 (2006): e473.

\bibitem{messina2014global} Messina, Jane P., et al. "A global compendium of human dengue virus occurrence." Scientific Data 1.1 (2014): 1-6.

\bibitem{pigott2014global} Pigott, David M., et al. "Global database of leishmaniasis occurrence locations, 1960–2012." Scientific Data 1.1 (2014): 1-7.

\bibitem{mylne2014comprehensive} Mylne, Adrian, et al. "A comprehensive database of the geographic spread of past human Ebola outbreaks." Scientific data 1.1 (2014): 1-10.

\bibitem{xu2020epidimiological} Xu, Bo, et al. "Epidemiological data from the COVID-19 outbreak, real-time case information." Scientific data 7.1 (2020): 1-6.

\bibitem{dong2020interactive} Dong, Ensheng, Hongru Du, and Lauren Gardner. "An interactive web-based dashboard to track COVID-19 in real time." The Lancet infectious diseases 20.5 (2020): 533-534.

\bibitem{wang2020covid} Wang, Lucy Lu, et al. "CORD-19: The Covid-19 Open Research Dataset." ArXiv (2020).

\bibitem{chen2020first} Chen, Emily, Kristina Lerman, and Emilio Ferrara. "Covid-19: The first public coronavirus twitter dataset." arXiv preprint arXiv:2003.07372 (2020).

\bibitem{covid2020india} COVID-19 India Org Data Operations Group (2020).


\bibitem{jain2010data} Jain, Anil K. "Data clustering: 50 years beyond K-means." Pattern recognition letters 31.8 (2010): 651-666.

\end{thebibliography}
